\documentclass[12pt]{article}

\usepackage{macros}

\usepackage{xcolor}

\usepackage{amsopn}

\usepackage{lmodern}
\usepackage{amsmath}

\usepackage{graphicx}

\usepackage{amssymb,amsmath}

\usepackage{relsize}

\usepackage{subfigure}

\usepackage{url}

\usepackage{graphicx}

\pdfoutput=1

\usepackage{amsmath,amssymb,color}

\def\x{\mathbf{x}}

\def\R{\mathbb{R}}

\def\X{\mathbf{X}}

\def\V{\mathbf{V}}
\def\I{\mathbf{I}}
\def\L{\mathcal{L}}
\def\hsig{\hat{\Sigma}}
\def\v{\mathbf{v}}
\def\pp{{p^\prime}}
\def\mmu{\boldsymbol{\mu}}

\usepackage{multicol}

\newcommand{\mcl}[2]{\multicolumn{#1}{c}{#2}}

\title{Optimal Projections for Gaussian Discriminants}

\begin{document}

\author{\name David P.\ Hofmeyr \hfill
{\small \textmd{ Department of Statistics and Actuarial Science}}\\
\name Francois Kamper \hfill {\small \textmd{Stellenbosch University}}\\
\name Michail C. Melonas \hfill {\small \textmd{7600, South Africa}}
}

\maketitle

\begin{abstract}
The problem of obtaining optimal projections for performing discriminant analysis with Gaussian class densities is studied. Unlike in most existing approaches to the problem, the focus of the optimisation is on the multinomial likelihood based on posterior probability estimates, which directly captures discriminability of classes. In addition to the more commonly considered problem, in this context, of classification, the unsupervised clustering counterpart is also considered. Finding optimal projections offers utility for dimension reduction and regularisation, as well as instructive visualisation for better model interpretability. Practical applications of the proposed approach show considerable promise for both classification and clustering. Code to implement the proposed method is available in the form of an {\tt R} package from \url{https://github.com/DavidHofmeyr/OPGD}.
\end{abstract}

\paragraph{Keywords:} Classification; clustering; dimension reduction; feature extraction; visualisation

\section{Introduction} \label{sec:intro}

In this paper we study the problem of finding optimal linear projections of a set of data, for the purpose of discriminant analysis. We study both the supervised classification problem, in which a set of known {\em class labels} is used to guide the selection of the projection to one which enhances the discriminability of the points in different classes from one another; and the unsupervised clustering problem, in which the relative spatial relationships between points are used to guide the projection to one on which subsets of points in the data (clusters) each appear more internally cohesive, while simultaneously (as entire groups) more separated from other cohesive groups. We focus specifically on the case in which the discriminability of classes/clusters is measured in terms of the multinomial likelihood based on estimates of the posterior probabilities of class/cluster membership. These posterior probability estimates arise from the standard application of Bayes' theorem, where we model each class/cluster conditional distribution with a multivariate Gaussian density.

The remaining paper is organised as follows. In Section~\ref{sec:background} we provide an explicit introduction to the problem of discriminant analysis for classification, as well as covering existing literature on the topic. In Section~\ref{sec:methodology} we elaborate on our adopted methodology and cover explicitly the practical issues associated with its implementation. In Section~\ref{sec:classificationexperiments} we discuss the results of experiments using our method when applied to publicly available data sets from diverse applications areas, and in comparison with existing methods from the literature. Then, in Section~\ref{sec:clustering}, we introduce a simple modification to our approach which allows us to apply it to the problem of clustering, and investigate the practical relevance of this modification for improving the performance of Gaussian Mixture Models. We then conclude our work in Section~\ref{sec:conclusions} with a discussion of our findings and experiences with the method.

\section{Discriminant Analysis} \label{sec:background}

Discriminant analysis in a probabilistic framework is concerned with estimating the probabilities $P(Y = k |\X = \x); k = 1, ..., K,$ via a simple reformulation based on Bayes' theorem, i.e.,
\begin{align}\label{eq:bayes}
    P(Y = k | \X = \x) &= \frac{\pi_kf_{X|Y=k}(\x)}{\sum_{l=1}^K \pi_lf_{X|Y=l}(\x)},
\end{align}
where we have used $f_{X|Y=k}$ to denote the density of the random variable $X | Y = k$, and $\pi_k$ is used to represent the {\em prior class probability} $P(Y = k)$. Different classification models in this framework differ in how they estimate the functions $f_{X|Y=k}$. it is almost universal that the prior probabilities are estimated as $\hat \pi_k = \frac{n_k}{n}; n_k:= \sum_{i=1}^n I(y_i=k)$, where $I$ is the indicator function and $\{y_1, ..., y_n\}$ are the labels associated with the data, $\{\x_1, ..., \x_n\}$. A very popular approach is to estimate the {\em posterior probabilities} in Eq.~(\ref{eq:bayes}) by replacing $f_{X|Y=k}$ with the Gaussian density with mean $ \mmu_k$ and covariance matrix $ \Sigma_k$, using suitably chosen estimates of these parameters, denoted by $\hat \mmu_k$ and $\hat \Sigma_k$. That is,
\begin{align}
\widehat{P(Y = k| \X = \x)} &= \frac{\hat \pi_k \phi_{\hat\mmu_k, \hat \Sigma_k}(\x)}{\sum_{l=1}^K \hat \pi_l \phi_{\hat\mmu_l, \hat \Sigma_l}(\x)} \label{eq:bayesGMM}\\
\phi_{\mmu, \Sigma}\left(\x\right) &:= \frac{1}{(2\pi)^{d/2}| \Sigma^{1/2}|}\exp\left(-\frac{1}{2}(\x - \mmu)^\top \Sigma^{-1}(\x - \mmu)\right).
\end{align}
Of these methods, Linear Discriminant Analysis~\citep[LDA]{citeLDA}, in which a common covariance matrix is used for all values of $k$; and Quadratic Discriminant Analysis~(QDA), in which a freely estimated covariance matrix is given separately to each class, are the most well known, and are often seen to represent opposite ends of a spectrum of Gaussian discriminant models of varying complexity. Intermediaries on this spectrum include, for example, regularised discriminant models, such as those described by~\cite{regularisedDA}; and models where classes share a common principal component structure for their covariance matrices~\citep{CPCDA}. Slightly tangential to this spectrum is the diagonal discriminant analysis model, in which each class is assumed to have a diagonal covariance matrix. 

The most flexible class of discriminant models places no restriction on the estimates for $f_{X|Y=k}$, and includes kernel density discriminant analysis~\citep{hand1982}, in which a full description of each density, $f_{X|Y=k}, k = 1, ..., K$, is obtained using kernel density estimation (or some other non-parametric method). A popular restriction to this class includes variants of the kernel Na{\"i}ve Bayes
classifier~\citep[NB]{john1995}, in which the estimates of $f_{X|Y=k}$ are assumed to factorise over their margins (i.e., the elements of $X|Y=k$ are treated as independent). Notice that the diagonal discriminant analysis model corresponds to the combination of QDA and the NB assumption. Other discriminant models which offer more flexibility than the class of Gaussian discriminants described in the previous paragraph include Mixture Discriminant Analysis~\citep[MDA]{MDA}, in which each class density is estimated using a simple mixture model; and Flexible Discriminant Analysis~\citep[FDA]{FDA}, where a Gaussian discriminant model is fit to a non-linear transformation of the data.


Of similar motivation to our approach, {\em discriminant feature extraction}~\citep{zhuHastie} is the process of identifying a collection of univariate projections of the data along which classes appear optimally discriminable, under a chosen model. By far the most well known approach of this type is the reduced rank Linear Discriminant Analysis approach. It is straightforward to show (see, for example, \cite{ESL}) that there exist at most $K-1$ univariate projections on which exact discrimination under the LDA model is possible. This comes from the fact that the $K$ class means lie in a subspace of dimension $K-1$, and since the sharing of a common covariance matrix means that a single scaling of the data projected into this $K-1$ dimensional subspace allows for exact discrimination based solely on the Euclidean distances between a point and the projected means, as well as the estimated prior probabilities. Due to this fact, and the ubiquity with which this approach is applied, the reduced rank version of LDA has become standard. It can also be shown that the discriminant features correspond with the leading eigenvectors of the matrix $\hat \Sigma_W^{-1}\hat \Sigma_B$, where $\hat \Sigma_W$ is the shared covariance matrix for the classes, and $\hat \Sigma_B = \frac{1}{n}\sum_{k=1}^K n_k (\hat \mmu_k - \bar \mmu)(\hat \mmu_k - \bar \mmu)^\top$ is the contribution to the total data covariance arising from the relative locations of the class means. Here $\bar \mmu = \frac{1}{n}\sum_{i=1}^n \x_i$ is the overall mean of the data. This is a convenient fact, as when fewer than $K-1$, say $\pp$, features are desired, one simply selects the first $\pp$ eigenvectors; these being those which contribute the most to the total discrimination of the classes. In the framework of quadratic discriminant analysis, Sliced Average Variance Estimation~\cite[SAVE]{save} has been successfully applied to obtain discriminant features. In this context the discriminant features are taken as the leading eigenvectors of the matrix $M = \sum_{k=1}^K \hat\pi_k (\I - \hat \Sigma_k)^2$, where here the entire data set is first {\em sphered}, i.e., transformed to jointly (combining all classes) have identity covariance matrix. Notice that, with the fact that the data have identity covariance, we have $\hat \Sigma_B + \hat \Sigma_W = \I$, and so $M = \sum_{k=1}^K \hat \pi_k \left(\hsig_B + \hsig_W - \hsig_k\right)^2$. ~\cite{cook2000identifying} have shown that the span of $M$ is equal to the span of the class mean vectors, $\hat \mu_1, ..., \hat \mu_K$, and the differences $\hsig_k - \hsig_{k-1}$ for $k = 2, ..., K$, which provides sufficient information for classification with QDA. The leading eigenvectors of $M$ are thus sensible candidates for discriminant features using the QDA model. Methods like LDA and SAVE are appealing for their fast computation, since their objectives are formulated as quadratic forms based on the class covariance matrices and the between class covariance, $\hat \Sigma_B$, and so fast eigen-solvers can be used to obtain optimal solutions.

More general frameworks, such as those adopted by~\cite{zhuHastie} and~\cite{calo2007gaussian}, use semi- or non-parametric estimates for the class conditional densities, and attempt to obtain discriminant features using a forward procedure based on projection pursuit. Specifically, a projection vector $\v$ is found by maximising the disciminatory capacity in the estimated class conditional densities, $\hat f_{\v^\top X|Y = 1}, ..., \hat f_{\v^\top X |Y = K}$. The data are then modified in such a way that they are no longer easily discriminable along the direction $\v$, and then another projection vector is obtained using the modified data. This process can be iterated in order to obtain the desired number of discriminant features. The reason for modifying the data before obtaining the next projection vector is to ensure that it is not simply re-discovering the discrimination of the classes already explained by the features obtained previously. The form of the density estimates $\hat f_{\v^\top X|Y = 1}, ..., \hat f_{\v^\top X |Y = K}$, as well as the measure used to determine the discriminatory capacity represented therein, is what separates different methods of this type. For example,~\cite{zhuHastie} use fully non-parametric density estimation, and maximise an objective based  on the Likelihood Ratio statistic for the test $H_0: f_{X|Y=k} = f_{X|Y=j} \ \forall j, k$, against $H_A: f_{X|Y=k} \not = f_{X|Y=j}$ for some $j, k$. Specifically, they focus on maximising over $\v \in \R^p; ||\v||=1$, the objective

\begin{align}\label{eq:zhu}
    LR(\v) = \prod_{i=1}^n\frac{\hat f_{\v^\top X|Y=y_i}(\v^\top\x_i)}{\hat f_{\v^\top X}(\v^\top\x_i)}.
\end{align}

Notice that reduced rank LDA, described above, can be formulated as a special case of this formulation, when each of the densities $f_{\v^\top X}$ is fit using a Gaussian density with mean $\v^\top\hat \mmu_k$ and variance $\v^\top \hat \Sigma_W \v$, and the null model is fit with a Gaussian density with mean $\v^\top\bar \mmu$, and variance $\v^\top \hat \Sigma_W \v$. For this reason the LDA discriminant features can also be seen as those which maximise the ANOVA statistic for the projected data~\citep{citeLDA}.
%

It is sensible to expect that by applying a discriminant rule based on Eq.~(\ref{eq:bayes}), applied to the data projected onto the features extracted by optimising~(\ref{eq:zhu}), one is likely to obtain reasonably accurate classification. However, if we examine the total discriminatory power which these features capture, as measured by the objective in~(\ref{eq:zhu}), we see that this approach sacrifices access to information available through interactions between these derived features during optimisation. Specifically, if $\v_1, ..., \v_\pp$ are the identified projections, and $\V \in \R^{p\times \pp}$ is the matrix with these features as columns, then in general,
\begin{align}\label{eq:biasZhu}
\prod_{i=1}^n \hat \pi_{y_i}\prod_{j=1}^\pp \frac{\hat f_{\v_j^\top X|Y = y_i}(\v_j^\top\x_i)}{\hat f_{\v_j^\top X}(\v_j^\top \x_i)} \not = \prod_{i=1}^n \frac{\hat \pi_{y_i}\hat f_{\V^\top X | Y=y_i}(\V^\top \x_i)}{\hat f_{\V^\top X}(\V^\top \x_i)}.
\end{align}
Notice that in relation to the context of Eq.~(\ref{eq:bayes}), the denominator terms above would be ideally captured by the densities $\hat f_{\v_j^\top X} = \sum_{k=1}^K \hat \pi_k \hat f_{\v_j^\top X | Y = k}$ and $\hat f_{\V^\top X} = \sum_{k=1}^K \hat \pi_k \hat f_{\V^\top X | Y = k}$. With a similar justification to that applied in the context of na{\"i}ve Bayes, it is quite reasonable to expect that features can be found along which the two numerator terms in~(\ref{eq:biasZhu}) are similar. However, it is unlikely that the null model, i.e., the denominator term in the likelihood ratio, represents a density which factorises over its margins, which would be necessary for~(\ref{eq:biasZhu}) to be a reasonable approximation. Indeed, applying a cursory thought to the problem, one finds that for a mixture of discriminable components to allow such factorisation, it would likely require the arrangement of the components to be fairly contrived; for example, lying on a precise grid or lattice.

\section{Optimal Projections for Gaussian Discriminants} \label{sec:methodology}

We study the problem of obtaining an optimal projection matrix, $\V \in \R^{p\times p'}$, for the purpose of discriminant analysis, in which each class conditional density is fit with a Gaussian. We use the word ``projection'' in a slightly more liberal sense than in some other contexts. In particular, we place no explicit restrictions, such as orthonormality, on the matrix $\V$, and so allow the magnitudes of the columns of $\V$ to also optimally accommodate 
scaling factors for the features of the projected data, $\X\V \in \R^{n\times \pp}$; where $\X \in \R^{n\times p}$ is the data matrix, with observations stored row-wise.
%
%
In the remainder we will occasionally refer to the projection of the data on $\V$, as well as to the projection of the data into the subspace defined by $\V$, which we mean to be interpreted practically equivalent.

The fundamental difference between ours and many existing approaches is that we focus on optimising the likelihood objective based on the original classification rule, i.e.,
\begin{align}\label{eq:classlik}
\L(\V) := \prod_{i=1}^n \frac{\hat \pi_{y_i}\hat f_{\V^\top X | Y=y_i}(\V^\top \x_i)}{\sum_{l=1}^K\hat \pi_{l}\hat f_{\V^\top X | Y=l}(\V^\top \x_i)}.
\end{align}
That is, the multinomial likelihood in which the probabilities associated with each observation are given by the posterior estimates arising from the standard application of Bayes' theorem in Eq.~(\ref{eq:bayes}). In the remainder we will refer to this as the {\em classification likelihood}.
Of the existing methods of which we are aware, only those of~\cite{peltonen2005discriminative} and \cite{peltonen} use the classification likelihood in order to optimise projections. Both of these estimate each class conditional density with a Gaussian mixture, as in MDA. It is surprising to us that this problem has apparently not been considered explicitly for the simpler Gaussian case. Although it is clear that this is a special case of these existing methods, their utilisation of mixtures allows them to avoid the practical difficulty of accommodating different covariance matrices for the components. 
Specifically, these methods circumvent this difficulty by either using a fixed isotropic covariance for each mixture component~\citep{peltonen2005discriminative} or by using an alternating optimisation procedure~\citep{peltonen} in which $\V$ is updated based on a gradient step in which the parameters of the mixture components within the projected densities are assumed constant, and a step in which these mixture parameters are updated for the new projection. This alternating procedure allows the authors to avoid having to obtain an exact expression for the gradient of the overall objective, since by ignoring the effect of varying $\V$ on the mixture parameters the problem is vastly simplified. However, the objectives used in the two alternating steps are not the same. In particular the mixture parameters are updated using the standard maximum likelihood objective for mixtures. As a result there is no guarantee that this alternating approach will lead to an increase in the objective of interest, i.e.,~(\ref{eq:classlik})\footnote{We note that it is not only pathological examples where such failure occurs, and we encountered this phenomenon frequently in experimentation with their approach even when each class density is fit with a single Gaussian component.}.

The approach which we adopt is to directly optimise~(\ref{eq:classlik}) using gradient based optimisation techniques. We give explicit details of the necessary derivations in the next subsection. 
%
%
Although this could be generalised\footnote{A combination of a simple modification of our derived gradients and the model formulation given by~\cite{peltonen} would allow considerable additional flexibility.}, for simplicity we focus here only on the case in which each $\hat f_{\V^\top X|Y=k}$ above is represented by a Gaussian density. We also add a simplifying (and regularising) restriction that, within the subspace defined by the projection on $\V$, the covariance matrix of each of the components is diagonal. Because we have freedom over the selection of $\V$, this is very similar to the restriction imposed by the common principal components model. 
%
%
The difference between our approach of optimising the classification likelihood and the common principal components model is similar to the difference between maximising the conditional likelihood of $\{y_1, ..., y_n\}$ given $\{\x_1, ..., \x_n\}$, as opposed to that of $\{\x_1, ..., \x_n\}$ given $\{y_1, ..., y_n\}$. That is, where the common principal components model focuses on how well the resulting covariance matrices for the classes best describe their associated observations, our objective is motivated by obtaining the transformation which best aligns the posterior probabilities for the observations with their observed class labels. 


\subsection{Optimising $\L(\V)$}

In order to obtain an optimal projection for classification under the Gaussian discriminant model, we focus on maximising the classification likelihood given in~(\ref{eq:classlik}), in which each $\hat f_{\V^\top X|Y=k}$ represents a Gaussian density with diagonal covariance matrix. As is common, we directly optimise the logarithm of this likelihood, which is thus given by
\begin{align*}
    \ell(\V) &= \sum_{i=1}^n \log\left(\frac{\hat \pi_{y_{i}} \phi_{\V^\top \hat\mmu_{y_i}, \Delta(\V^\top\hat \Sigma_{y_i}\V)}(\V^\top \x_i)}{\sum_{l=1}^K \hat \pi_l \phi_{\V^\top \hat\mmu_l, \Delta(\V^\top \hat \Sigma_l \V)}(\V^\top \x_i)}\right)\\
    &= \sum_{i=1}^n\left( \log\left(\hat \pi_{y_{i}} \phi_{\V^\top \hat\mmu_{y_i}, \Delta(\V^\top\hat \Sigma_{y_i}\V)}(\V^\top \x_i)\right) - \log\left(\sum_{l=1}^K \hat \pi_l \phi_{\V^\top \hat\mmu_l, \Delta(\V^\top \hat \Sigma_l \V)}(\V^\top \x_i)\right)\right),
\end{align*}
where for a square matrix $A$ we use $\Delta (A)$ to be the diagonal matrix with $\Delta(A)_{ii} = A_{ii}$ and $\Delta(A)_{ij} = 0$ for $i \not = j$. By considering the effect of $\V$ on the parameters of each component, i.e., $\V^\top \mmu_k$ and $\Delta(\V^\top \hat \Sigma_k \V)$, during determination of the gradient of $\ell(\V)$, we are able to directly apply gradient ascent on this objective. For convenience in the following derivations,
we use the notation $\phi^{(k)}_\V$  for the density $\phi_{\V^\top\hat\mmu_{k}, \Delta(\V^\top\hat \Sigma_{k}\V)}$, for $k=1,2,\hdots,K$. 
The log-likelihood then becomes
\begin{align*}
    \ell(\V) &= \sum_{i=1}^{n}\text{log}(\hat{\pi}_{y_{i}}) +\sum_{i=1}^n  \log\left(  \phi^{(y_{i})}_{\mathbf{V}}(\mathbf{V}^\top\mathbf{x}_{i})\right)  - \sum_{i=1}^{n}\log\left(\sum_{l=1}^K \hat \pi_l \phi^{(l)}_{\mathbf{V}}(\mathbf{V}^\top\mathbf{x}_{i})\right) \\
    &= \sum_{i=1}^{n}\text{log}(\hat{\pi}_{y_{i}}) + \ell_{1}(\V) - \ell_{2}(\V),
\end{align*}
where $\ell_{1}(\V) = \sum_{i=1}^n \log\left( \phi^{(y_{i})}_{\mathbf{V}}(\mathbf{V}^\top\mathbf{x}_{i})\right) $ and $\ell_{2}(\V) = \sum_{i=1}^{n}\log\left(\sum_{l=1}^K \hat \pi_l \phi^{(l)}_{\mathbf{V}}(\mathbf{V}^\top\mathbf{x}_{i})\right)$.
\\
\indent First we show that $\ell_{1}(\mathbf{V}) =  c - \frac{1}{2}\sum_{k=1}^{K}n_{k}\sum_{t=1}^{p'}\text{log}(\mathbf{v}_{t}^\top\hat{\Sigma}_{k}\mathbf{v}_{t})$, where $c$ does not depend on $\mathbf{V}$. Consider:
\begin{align}
\ell_{1}(\mathbf{V}) &= \sum_{k=1}^{K}\sum_{y_{i}=k} \text{log}(\phi^{(k)}_{\mathbf{V}}(\mathbf{V}^\top\mathbf{x}_{i})) \notag \\
&= -\frac{np'\text{log}(2 \pi)}{2} - \frac{1}{2}\sum_{k=1}^{K}n_{k}
\text{log}|\Delta(\V^\top\hat \Sigma_{k}\V)| \notag \\
&- \frac{1}{2}\sum_{k=1}^{K}\sum_{y_{i}=k}(\mathbf{V}^\top\mathbf{x}_{i}-\mathbf{V}^\top\hat{\boldsymbol{\mu}}_{k})^\top(\Delta(\V^\top\hat \Sigma_{k}\V))^{-1}(\mathbf{V}^\top\mathbf{x}_{i}- \mathbf{V}^\top\hat{\boldsymbol{\mu}}_{k}). \notag 
\end{align}
Note that
\begin{align}
 &(\mathbf{V}^\top\mathbf{x}_{i}-\mathbf{V}^\top\hat{\boldsymbol{\mu}}_{k})^\top(\Delta(\V^\top\hat \Sigma_{k}\V))^{-1}(\mathbf{V}^\top\mathbf{x}_{i}- \mathbf{V}^\top\hat{\boldsymbol{\mu}}_{k}) \notag \\ 
 =& (\mathbf{x}_{i}-\hat{\boldsymbol{\mu}}_{k})^\top\mathbf{V}(\Delta(\V^\top\hat \Sigma_{k}\V))^{-1}\mathbf{V}^\top(\mathbf{x}_{i}- \hat{\boldsymbol{\mu}}_{k}) \notag \\
 =& \sum_{t=1}^{p'}\frac{\mathbf{v}_{t}^\top(\mathbf{x}_{i}-\hat{\boldsymbol{\mu}}_{k})(\mathbf{x}_{i}-\hat{\boldsymbol{\mu}}_{k})^\top\mathbf{v}_{t}}{\mathbf{v}_{t}^\top\hat{\Sigma}_{k}\mathbf{v}_{t}}.
\end{align}
The result follows from 
\begin{align*}
\sum_{k=1}^{K}&\sum_{y_{i}=k}(\mathbf{V}^\top\mathbf{x}_{i}-\mathbf{V}^\top\hat{\boldsymbol{\mu}}_{k})^\top(\Delta(\V^\top\hat \Sigma_{k}\V))^{-1}(\mathbf{V}^\top\mathbf{x}_{i}- \mathbf{V}^\top\hat{\boldsymbol{\mu}}_{k}) \notag \\ 
&=\sum_{k=1}^{K}\sum_{y_{i}=k}\sum_{t=1}^{p'}\frac{\mathbf{v}_{t}^\top(\mathbf{x}_{i}-\hat{\boldsymbol{\mu}}_{k})(\mathbf{x}_{i}-\hat{\boldsymbol{\mu}}_{k})^\top\mathbf{v}_{t}}{\mathbf{v}_{t}^\top\hat{\Sigma}_{k}\mathbf{v}_{t}}
=\sum_{k=1}^{K}\sum_{t=1}^{p'}\sum_{y_{i}=k}\frac{\mathbf{v}_{t}^\top(\mathbf{x}_{i}-\hat{\boldsymbol{\mu}}_{k})(\mathbf{x}_{i}-\hat{\boldsymbol{\mu}}_{k})^\top\mathbf{v}_{t}}{\mathbf{v}_{t}^\top\hat{\Sigma}_{k}\mathbf{v}_{t}}\\
&= \sum_{k=1}^{K}\sum_{t=1}^{p'}\frac{1}{\mathbf{v}_{t}^\top\hat{\Sigma}_{k}\mathbf{v}_{t}}
\mathbf{v}_{t}^\top\bigg{(}\sum_{y_{i}=k}(\mathbf{x}_{i}-\hat{\boldsymbol{\mu}}_{k})(\mathbf{x}_{i}-\hat{\boldsymbol{\mu}}_{k})^\top\bigg{)}\mathbf{v}_{t} 
= \sum_{k=1}^{K}\sum_{t=1}^{p'}\frac{n_{k} \mathbf{v}_{t}^\top\hat{\Sigma}_{k}\mathbf{v}_{t}}{\mathbf{v}_{t}^\top\hat{\Sigma}_{k}\mathbf{v}_{t}}
 \\
&= \sum_{k=1}^{K}\sum_{t=1}^{p'}n_{k} 
= p'n,
\end{align*}
and $|\Delta(\V^\top\hat \Sigma_{k}\V)| = \prod_{t=1}^{p'}\mathbf{v}_{t}^\top\hat{\Sigma}_{k}\mathbf{v}_{t}$.
The gradient of $\ell_{1}$ with respect to the $j$-th column of $\V$, $\mathbf{v}_{j}$, is therefore
\begin{equation}
\frac{\partial }{\partial \mathbf{v}_{j}} \ell_{1}(\V) = -\bigg{(}\sum_{k=1}^{K}\frac{n_{k}}{\mathbf{v}_{j}'\hat{\Sigma}_{k}\mathbf{v}_{j}} \hat{\Sigma}_{k}\bigg{)}\mathbf{v}_{j}.    
\end{equation}
For the differentiation of $\ell_{2}$ with respect to $\mathbf{v}_{j}$ we note that
\begin{equation}
\frac{\partial }{\partial \mathbf{v}_{j}} \ell_{2}(\V) = \sum_{i=1}^{n}\bigg{(}\frac{1}{\sum_{l=1}^K \hat \pi_l \phi^{(l)}_{\mathbf{V}}(\mathbf{V}^\top\mathbf{x}_{i})}\sum_{k=1}^{K}\hat{\pi}_{k}\frac{\partial}{\partial \mathbf{v}_{j}}\phi^{(k)}_{\mathbf{V}}(\mathbf{V}^\top\mathbf{x}_{i}) \bigg{)}.   \label{LS17.5}
\end{equation}
Now consider that
\begin{align}
\frac{\partial}{\partial \mathbf{v}_{j}} & \phi^{(k)}_{\mathbf{V}}(\mathbf{V}^\top\mathbf{x}_{i}) \notag \\
=& (2\pi)^{-\frac{p'}{2}} \frac{\partial}{\partial \mathbf{v}_{j}} |\Delta(\V^\top\hat \Sigma_{k}\V)|^{-\frac{1}{2}}\text{exp}\bigg{(}-\frac{1}{2}\sum_{t=1}^{p'}\frac{\mathbf{v}_{t}^\top(\mathbf{x}_{i}-\hat{\boldsymbol{\mu}}_{k})(\mathbf{x}_{i}-\hat{\boldsymbol{\mu}}_{k})^\top\mathbf{v}_{t}}{\mathbf{v}_{t}^\top\hat{\Sigma}_{k}\mathbf{v}_{t}}\bigg{)} \notag \\
=& (2\pi)^{-\frac{p'}{2}} \text{exp}\bigg{(}-\frac{1}{2}\sum_{t=1}^{p'}\frac{\mathbf{v}_{t}^\top(\mathbf{x}_{i}-\hat{\boldsymbol{\mu}}_{k})(\mathbf{x}_{i}-\hat{\boldsymbol{\mu}}_{k})^\top\mathbf{v}_{t}}{\mathbf{v}_{t}^\top\hat{\Sigma}_{k}\mathbf{v}_{t}}\bigg{)}\frac{\partial}{\partial \mathbf{v}_{j}} |\Delta(\V^\top\hat \Sigma_{k}\V)|^{-\frac{1}{2}} \notag \\
&+  (2\pi)^{-\frac{p'}{2}}  |\Delta(\V^\top\hat \Sigma_{k}\V)|^{-\frac{1}{2}} \frac{\partial}{\partial \mathbf{v}_{j}} \text{exp}\bigg{(}-\frac{1}{2}\sum_{t=1}^{p'}\frac{\mathbf{v}_{t}^\top(\mathbf{x}_{i}-\hat{\boldsymbol{\mu}}_{k})(\mathbf{x}_{i}-\hat{\boldsymbol{\mu}}_{k})^\top\mathbf{v}_{t}}{\mathbf{v}_{t}^\top\hat{\Sigma}_{k}\mathbf{v}_{t}}\bigg{)}. \label{LS17.1}
\end{align}
Next we evaluate the gradient of $|\Delta(\V^\top\hat \Sigma_{k}\V)|^{-\frac{1}{2}} = \prod_{t=1}^{p'}\frac{1}{\sqrt{\mathbf{v}_{t}^\top\hat{\Sigma}_{k}\mathbf{v}_{t}}}$ with 
respect to $\mathbf{v}_{j}$. Specifically,
\begin{align}
\frac{\partial}{\partial \mathbf{v}_{j}}\prod_{t=1}^{p'}\frac{1}{\sqrt{\mathbf{v}_{t}^\top\hat{\Sigma}_{k}\mathbf{v}_{t}}} &= \bigg{(}\prod_{t\neq j}\frac{1}{\sqrt{\mathbf{v}_{t}^\top\hat{\Sigma}_{k}\mathbf{v}_{t}}} \bigg{)}
\frac{\partial}{\partial \mathbf{v}_{j}} \frac{1}{\sqrt{\mathbf{v}_{j}^\top\hat{\Sigma}_{k}\mathbf{v}_{j}}} \notag = -\bigg{(}\prod_{t\neq j}\frac{1}{\sqrt{\mathbf{v}_{t}^\top\hat{\Sigma}_{k}\mathbf{v}_{t}}} \bigg{)}
\frac{\hat{\Sigma}_{k}\mathbf{v}_{j}}{(\mathbf{v}_{j}^\top\hat{\Sigma}_{k}\mathbf{v}_{j})^{\frac{3}{2}} }
 \notag \\
&= -\bigg{(}\prod_{t=1}^{p'}\frac{1}{\sqrt{\mathbf{v}_{t}^\top\hat{\Sigma}_{k}\mathbf{v}_{t}}} \bigg{)}
\frac{\hat{\Sigma}_{k}\mathbf{v}_{j} }{\mathbf{v}_{j}^\top\hat{\Sigma}_{k}\mathbf{v}_{j}}
= -\frac{|\Delta(\V^\top\hat \Sigma_{k}\V)|^{-\frac{1}{2}}}{\mathbf{v}_{j}^\top\hat{\Sigma}_{k}\mathbf{v}_{j}}\hat{\Sigma}_{k}\mathbf{v}_{j}. \label{LS17.2}
\end{align}
To complete the gradient in (\ref{LS17.1}) we need to evaluate
\begin{align}
\frac{\partial}{\partial \mathbf{v}_{j}} &\text{exp}\bigg{(}-\frac{1}{2}\sum_{t=1}^{p'}\frac{\mathbf{v}_{t}^\top(\mathbf{x}_{i}-\hat{\boldsymbol{\mu}}_{k})(\mathbf{x}_{i}-\hat{\boldsymbol{\mu}}_{k})^\top\mathbf{v}_{t}}{\mathbf{v}_{t}^\top\hat{\Sigma}_{k}\mathbf{v}_{t}}\bigg{)} \notag \\  
=&-\frac{1}{2}\text{exp}\bigg{(}-\frac{1}{2}\sum_{t=1}^{p'}\frac{\mathbf{v}_{t}^\top(\mathbf{x}_{i}-\hat{\boldsymbol{\mu}}_{k})(\mathbf{x}_{i}-\hat{\boldsymbol{\mu}}_{k})^\top\mathbf{v}_{t}}{\mathbf{v}_{t}^\top\hat{\Sigma}_{k}\mathbf{v}_{t}}\bigg{)} \frac{\partial}{\partial \mathbf{v}_{j}} \frac{\mathbf{v}_{j}^\top(\mathbf{x}_{i}-\hat{\boldsymbol{\mu}}_{k})(\mathbf{x}_{i}-\hat{\boldsymbol{\mu}}_{k})^\top\mathbf{v}_{j}}{\mathbf{v}_{j}^\top\hat{\Sigma}_{k}\mathbf{v}_{j}}. \notag \\
=& -\text{exp}\bigg{(}-\frac{1}{2}\sum_{t=1}^{p'}\frac{\mathbf{v}_{t}^\top(\mathbf{x}_{i}-\hat{\boldsymbol{\mu}}_{k})(\mathbf{x}_{i}-\hat{\boldsymbol{\mu}}_{k})^\top\mathbf{v}_{t}}{\mathbf{v}_{t}^\top\hat{\Sigma}_{k}\mathbf{v}_{t}}\bigg{)} \notag \\
&\times \bigg{(} (\mathbf{x}_{i}-\hat{\boldsymbol{\mu}}_{k})(\mathbf{x}_{i}-\hat{\boldsymbol{\mu}}_{k})^\top - \frac{\mathbf{v}_{j}^\top(\mathbf{x}_{i}-\hat{\boldsymbol{\mu}}_{k})(\mathbf{x}_{i}-\hat{\boldsymbol{\mu}}_{k})^\top \mathbf{v}_{j}}{\mathbf{v}_{j}^\top\hat{\Sigma}_{k}\mathbf{v}_{j}}\hat{\Sigma}_{k} \bigg{)}\frac{\mathbf{v}_{j}}{\mathbf{v}_{j}^\top\hat{\Sigma}_{k}\mathbf{v}_{j}}. \label{LS17.3}
\end{align}
Substitution of (\ref{LS17.2}) and (\ref{LS17.3}) into (\ref{LS17.1}) yields
\begin{align}
\frac{\partial}{\partial \mathbf{v}_{j}}\phi^{(k)}_{\mathbf{V}}(\mathbf{V}^\top\mathbf{x}_{i}) =& \frac{\phi^{(k)}_{\mathbf{V}}(\mathbf{V}^\top\mathbf{x}_{i})}{\mathbf{v}_{j}^\top \hat{\Sigma}_{k}\mathbf{v}_{j}} \notag \\
&\times \bigg{(} \bigg{(}\frac{\mathbf{v}_{j}^\top (\mathbf{x}_{i}-\hat{\boldsymbol{\mu}}_{k})(\mathbf{x}_{i}-\hat{\boldsymbol{\mu}}_{k})^\top\mathbf{v}_{j}}{\mathbf{v}_{j}^\top \hat{\Sigma}_{k}\mathbf{v}_{j}} - 1 \bigg{)}\hat{\Sigma}_{k}
- (\mathbf{x}_{i}-\hat{\boldsymbol{\mu}}_{k})(\mathbf{x}_{i}-\hat{\boldsymbol{\mu}}_{k})^\top \bigg{)} \mathbf{v}_{j}. \label{LS17.4}
\end{align}
Now let $p_{ik}(\mathbf{V}) = \frac{\hat{\pi}_{k}\phi^{(k)}_{\mathbf{V}}(\mathbf{V}^\top\mathbf{x}_{i})}{\sum_{l=1}^{K}\hat{\pi}_{l}\phi^{(l)}_{\mathbf{V}}(\mathbf{V}^\top\mathbf{x}_{i})}$ and $\alpha_{ik}(\mathbf{v}_{j}) = \frac{\mathbf{v}_{j}^\top (\mathbf{x}_{i}-\hat{\boldsymbol{\mu}}_{k})(\mathbf{x}_{i}-\hat{\boldsymbol{\mu}}_{k})^\top\mathbf{v}_{j}}{\mathbf{v}_{j}^\top \hat{\Sigma}_{k}\mathbf{v}_{j}}$, so that substitution of (\ref{LS17.4}) into (\ref{LS17.5}) gives
\begin{align}
\frac{\partial \ell_{2}}{\partial \mathbf{v}_{j}} &= \bigg{(}\sum_{i=1}^{n}\sum_{k=1}^{K} \frac{p_{ik}(\mathbf{V})}{\mathbf{v}_{j}^\top \hat{\Sigma}_{k}\mathbf{v}_{j}}
\big{(} (\alpha_{ik}(\mathbf{v}_{j}) - 1 )\hat{\Sigma}_{k}
- (\mathbf{x}_{i}-\hat{\boldsymbol{\mu}}_{k})(\mathbf{x}_{i}-\hat{\boldsymbol{\mu}}_{k})^\top \big{)}
\bigg{)}\mathbf{v}_{j}.
\end{align}
Finally, setting $\mathbf{S}_{k}(\mathbf{V}) = \sum_{i=1}^{n}p_{ik}(\mathbf{V})(\mathbf{x}_{i}-\hat{\boldsymbol{\mu}}_{k})(\mathbf{x}_{i}-\hat{\boldsymbol{\mu}}_{k})^\top$, we find that
\begin{align}
&\sum_{i=1}^{n}\sum_{k=1}^{K} \frac{p_{ik}(\mathbf{V})}{\mathbf{v}_{j}^\top \hat{\Sigma}_{k}\mathbf{v}_{j}}
\big{(} (\alpha_{ik}(\mathbf{v}_{j}) - 1 )\hat{\Sigma}_{k} 
- (\mathbf{x}_{i}-\hat{\boldsymbol{\mu}}_{k})(\mathbf{x}_{i}-\hat{\boldsymbol{\mu}}_{k})^\top \big{)} \notag \\    
=& \sum_{k=1}^{K} \frac{1}{\mathbf{v}_{j}^\top \hat{\Sigma}_{k}\mathbf{v}_{j}}\bigg{(}\sum_{i=1}^{n}p_{ik}(\mathbf{V})\alpha_{ik}(\mathbf{v}_{j}) -\sum_{i=1}^{n}p_{ik}(\mathbf{V}) \bigg{)}\hat{\Sigma}_{k} \notag \\
&- \sum_{k=1}^{K} \frac{1}{\mathbf{v}_{j}^\top \hat{\Sigma}_{k}\mathbf{v}_{j}}\sum_{i=1}^{n}p_{ik}(\mathbf{V})(\mathbf{x}_{i}-\hat{\boldsymbol{\mu}}_{k})(\mathbf{x}_{i}-\hat{\boldsymbol{\mu}}_{k})^\top \notag \\
=& \sum_{k=1}^{K} \frac{1}{\mathbf{v}_{j}^\top \hat{\Sigma}_{k}\mathbf{v}_{j}}\bigg{(} \frac{\mathbf{v}_{j}^\top\mathbf{S}_{k}(\mathbf{V})\mathbf{v}_{j}}{\mathbf{v}_{j}^\top \hat{\Sigma}_{k}\mathbf{v}_{j}} -\sum_{i=1}^{n}p_{ik}(\mathbf{V}) \bigg{)}\hat{\Sigma}_{k} 
- \sum_{k=1}^{K} \frac{1}{\mathbf{v}_{j}^\top \hat{\Sigma}_{k}\mathbf{v}_{j}}\mathbf{S}_{k}(\mathbf{V}).
\end{align}

\subsection{The Problem of Non-concavity, and Initialisation of $\V$}
The objective $\ell(\mathbf{V})$ is a non-concave function of 
$\mathbf{V}$, and hence the performance of our method will depend on the intialisation of $\mathbf{V}$. We have found that obtaining a warm start is frequently preferable to multiple random initialisations. This especially because the complexity of the gradient evaluations discussed above means that the practical running time of the method is considerably slower than competing approaches such as LDA and SAVE, and hence only few initialisations may frequently be reasonable. In practice, we found that taking the leading $p'$ eigenvalues of the matrix
\begin{equation}
\hat \Sigma_W^{-1}\hat \Sigma_B + \epsilon \hat{\Sigma}, \label{eq::ini1}
\end{equation}
where $\hat{\Sigma} = \frac{1}{n}\sum_{i=1}^{n}(\mathbf{x}_{i} - \bar{\boldsymbol{\mu}})(\mathbf{x}_{i} - \bar{\boldsymbol{\mu}})^\top$ is the covariance matrix of all of the data and $\epsilon$ is a small positive constant, frequently yields strong performance. Notice that the first term in the above is the matrix used to obtain discriminant features in LDA, and hence for small $\epsilon$ the leading $K-1$ eigenvectors will be very similar to the LDA features. Since this matrix has rank at mosk $K-1$, however, it is necessary to modify the objective so that more than $K-1$ dimensions can be obtained. The second term being proportional to the total data covariance means that these extra columns in the initialisation of $\V$ will be similar to the principal components within the null space of the leading $K-1$. In practice, to ensure a solution exists, we also add a small ridge to $\hat \Sigma_W$.
\indent


\subsection{Ordering the Columns of $\V$}
A limitation of our approach is that, unlike in the case of LDA and SAVE, whose final solutions are given by the eigenvectors of a matrix, there is not a natural ordering on the columns of $\V$. This is of primary importance in the context of visualisation, since it is necessary to produce instructive scatter plots based on the discriminant features. The first visualisation should be given by the features which contribute the most to the discrimination of the classes, and only if necessary will additional visualisations be sought.

We use a simple greedy approach to order the columns in the final solution. Specifically, we evaluate the classification likelihood for the univariate projections given by each of the columns of $\V$. The greatest likelihood determines the leading column of $\V$. We then iteratively augment the current set of ordered columns with each of the columns not yet utilised, and add that which yields the highest likelihood when added to the set already in place.


\section{Classification Experiments} \label{sec:classificationexperiments}
In this section an empirical comparison of our aproach, which we henceforth refer to as OPGD (Optimal Projection for Gaussian Discriminants), to
popular existing discriminant models based on Gaussian class densities is conducted. Depending on the size of the data set, we use either cross-validation or a standard ``training/validation'' split of the training data to select appropriate hyper-parameters. Performance is then compared based on prediction accuracy on an independent test set.
The methods included, and their respective ranges of hyper-parameters, are
\begin{enumerate}
    \item (Reduced rank) LDA, with selection of the number of discriminant features from 1 to $K-1$. We used the implementation in the {\tt R} package {\tt MASS}~\citep{MASSpackage}.
    \item Regularised Discriminant Analysis (RDA), in which the class covariance matrices are estimated by $\hat \Sigma_k = \alpha \tilde \Sigma_k + (1-\alpha) \hat \Sigma_W$. Here we have used $\tilde \Sigma_k$ to be the maximum likelihood estimate for the $k$-th class and $\hat{\Sigma}_W$ is the pooled estimate of the class covariance, as used in LDA. The value of $\alpha$ was selected from the set $\{0, \frac{1}{p-1}, \frac{2}{p-1}, ...,\frac{p-2}{p-1}, 1\}$, where $p$ is the dimension of the input space. Note that when $\alpha = 1$ this corresponds with the standard QDA model, and with $\alpha = 0$ this is equivalent to LDA.
    \item SAVE, with number of discriminant features selected from between 1 and $p$.
    \item Discriminant analysis with the Common Principal Components model (CPCDA), with the number of discriminant features (rank of the covariance matrices) selected from between 1 and $p$. We used the implementation in the {\tt R} package {\tt multigroup}~\citep{multigrouppackage} to obtain the estimate of the common principal components and corresponding scaling factors.
    \item The proposed method (OPGD) with the number of features (columns of $\mathbf{V}$) selected from between 1 and $p$. 
\end{enumerate}
All data sets under consideration are available from the UCI machine learning repository~\citep{UCI}. These are 
\begin{enumerate}
 \item
Vowel recognition.
\item
Statlog (landsat satellite).
\item
Image segmentation.
\item
Optical recognition of handwritten digits.
\end{enumerate}
We discuss the specifics for each application explicitly below.

\subsection{Vowel Recognition}
Here the objective is the recognition of vowel sounds from utterances taken from multiple speakers. There are 11 vowel sounds and 15 speakers. Each speaker spoke each vowel sound 6 times giving 990 utterances in total. The speech signals of 
each utterance have been preprocessed into 10 values, yielding
an input space of 10 dimensions. We follow the same experimental design as described in~\cite{ESL}, in which four males and four 
females are chosen for training (528 utterances) and four males and three females for testing (462 utterances).

A cross-validation procedure is applied to the training data in order to select the hyper-parameters of the methods under consideration. Eight folds are used and each fold corresponds to a different 
speaker. The hyper-parameters chosen are those which yield the smallest cross-validation error (misclassification error is used). Each model is then retrained on the complete training
data set using its chosen hyper-parameter.

The results are summarised in Table \ref{tab:vowel}. OPGD provides the smallest test error, 
followed by RDA and then LDA. SAVE and CPCDA performed 
relatively poorly.

\begin{table}
    \centering
    \begin{tabular}{c| c c c c c}
         & OPGD & LDA & RDA & SAVE & CPCDA\\
        \hline
        Test error & 0.4480519& 0.491342& 0.4718615&  0.5281385& 0.5649351 \\
        Hyper-parameter& 3 & 2 & 0.4444444& 10& 7 
    \end{tabular}
        \caption{Test errors and chosen hyper-parameters of the cross-validation procedure on the vowel recognition data.}
    \label{tab:vowel}
\end{table}

\subsection{Statlog}
The data consist of multi-spectral values of pixels in $3 \times 3$ neighbourhoods of a satellite image. Each pixel is represented by 4 multi-spectral values yielding an input
space of dimension 36. The objective is to classify the 
multi-spectral values of each neighbourhood into one of 7
classes. The data set consist of 6435 observations. It should be noted that only 6 of the 7 classes are present in the data (the ``mixture'' class is not present).

A split of 75\% training and 25\% testing was use. A third of the training data were used for validation, in order to select hyper-parameters, leading to an overall 50/25/25 training/validation/test split. 
%
Once hyper-parameters have been selected, the final models are trained on the entire training set. For our method, we initialise the optimisation of the final model with the solution which gives the lowest validation error to exploit the computations already performed.
Finally, the models are then assessed 
on the test set using misclassification error.

The entirety of the above procedure was performed 20 times, each corresponding to a different randomly generated training/validation/test split. 
%
The test errors corresponding to the different methods 
are summarized in the form of boxplots in the top-left graph of Figure \ref{class::figs}. We see that OPGD outperforms others by a considerable margin. RDA and SAVE yield
similar performance to one another, both significantly outperfroming LDA and CPCDA.

\subsection{Image Segmentation}
Each input observation consists of 19 values derived from 
a $3 \times 3$ neighbourhood of an image. These neighbourhoods were drawn randomly from a data base of 7 images and each neighbourhood corresponds to one of 7 classes. The objective is to classify the input observations into one of these 7 classes. \\
\noindent
\indent The region-pixel-count input is removed from the analysis since it is constant over all the observations.
The CPCDA method is particularly sensitive to the short-line-density-5 and short-line-density-2 inputs, and typically fails to provide output. We added small Gaussian perturbations to these data so that all models could be fit. These do not affect the output of the other methods appreciably.

\begin{figure}
    \centering
    \includegraphics[scale=0.7]{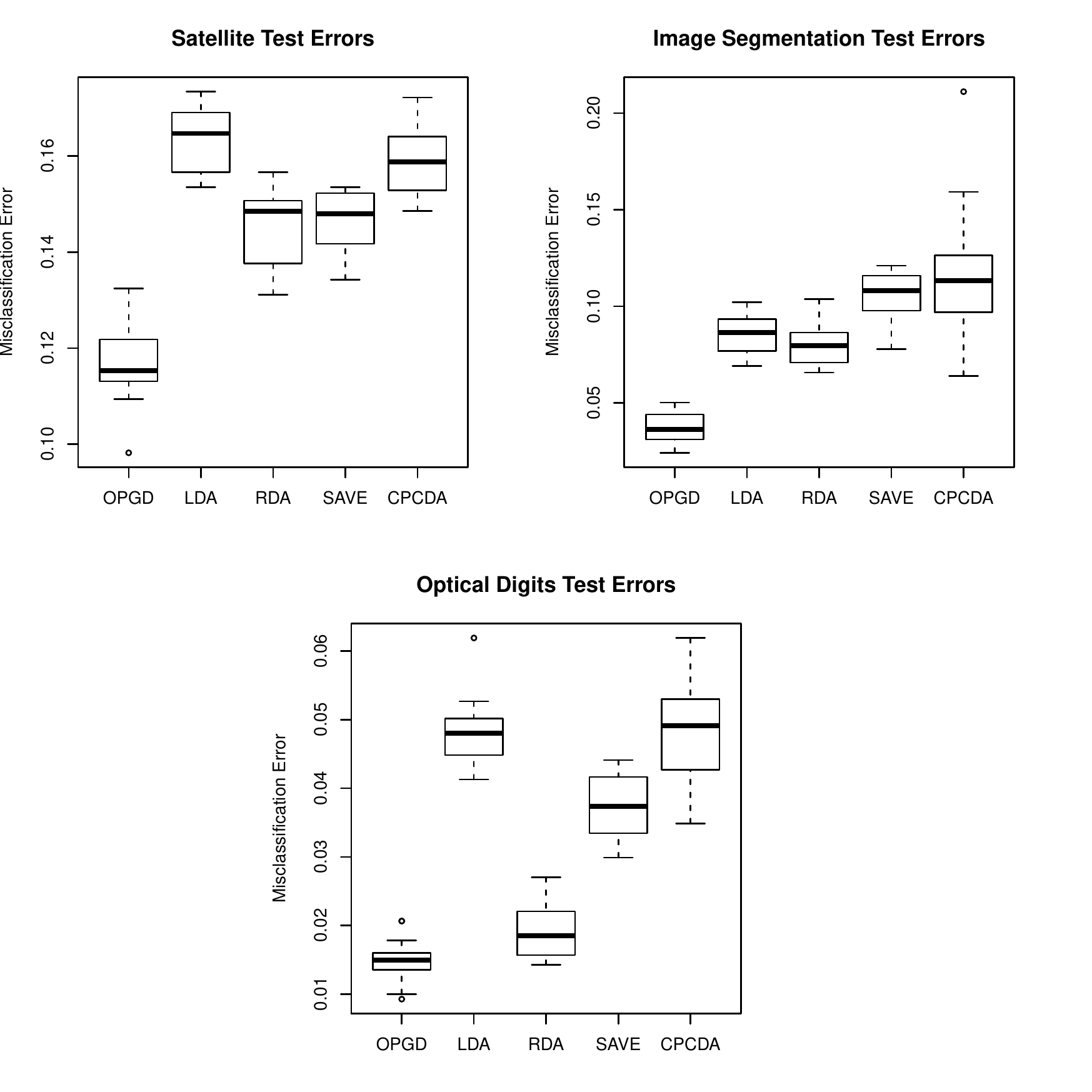}
    \caption{Results of the 20 training/validation/test splits applied to the statlog, image segmentation and
    optimal digit recognition data sets. The boxplots represent the test errors obtained over the different splits. All training/validation/test splits were done according to the 50/25/25 ratio.}
    \label{class::figs}
\end{figure}

The same proportions of training/validation/test as in the statlog data were applied. Again, 20 different random splits were generated
and the test errors recorded. The results are displayed in 
the top-right graph of Figure \ref{class::figs}.
Once more OPGD obtains encouraging performance in terms of classification accuracy. In this case LDA and RDA are similar in performance to one another. SAVE and CPCDA have similar average performance to one another, but with the performance of CPCDA model is considerably more variable. A possible reason for this is its sensitivity to the two short-line-density input variables discussed previously.

\subsection{Optical Recognition of Handwritten Digits}

In the final classification experiment we consider a data set containing images of handwritten digits, compressed to $8\times 8$ pixels, yielding 64 dimensions. The objective is to classify an image to one of the digits
between 0 and 9 (10 classes in total). 

We repeat the approach used for training and assessing models in the statlog and image segmentation data. The first and fortieth input variables were excluded since they were constant over all observations. To the remainder of the variables, independent Gaussian perturbations were added. This was done in order to obtain output for the CPCDA model.
Both OPGD an RDA obtain substantially better accuracy than the other three models considered, with OPGD itself being the most accurate.

\subsection{Remarks}
OPGD was compared to LDA, RDA, SAVE and CPCDA based on
four different data sets. In terms of test performance,
the results are encouraging, and OPGD 
outperformed its competitors in all applications. This
improvement in test error does not come without a cost, however, since the computation requirements of OPGD are considerably more demanding than these competing methods.
Everything considered, we believe that the potential for considerably improved accuracy in classification compared with other popular Gaussian discriminant models justifies the consideration of OPGD as an alternative to popular alternatives for problems of moderate size (up to thousands of points in tens of dimensions, or fewer), and that
methods of accelerating the algorithm should be considered in further research.

\section{Optimal Projections for Clustering with Gaussian Mixtures} \label{sec:clustering}

In this section we perform a preliminary exploration into the potential of the proposed method to enhance the clustering capabilities of Gaussian Mixture Models (GMMs). We use a slight modification of the objective which is better suited to the clustering context; i.e., where the labels are unknown. Specifically, suppose we have an initial estimate for the parameters of a GMM, denoted by $\hat \pi_1, ..., \hat \pi_K$; $\hat \mmu_1, ..., \hat \mmu_K$; and $\hsig_1, ..., \hsig_K$. We then focus on maximising the objective,
\begin{align*}
    \sum_{i=1}^n \log\left(\frac{\max_{l \in \{1, ..., K\}} \hat \pi_l \phi_{\V^\top \hat\mmu_l, \Delta(\V^\top\hat \Sigma_l\V)}(\V^\top \x_i)}{\sum_{l=1}^K \hat \pi_l \phi_{\V^\top \hat\mmu_l, \Delta(\V^\top \hat \Sigma_l \V)}(\V^\top \x_i)}\right) - \lambda ||\V^\top \V - \I||_F^2,
\end{align*}
where the difference from the objective used in the classification context is that the numerator terms are now taken as the maxima over the components. This effectively allows the points' assignments to the different components of the GMM to change during optimisation; which is appropriate given the unsupervised context. We have also added the term $-\lambda ||\V^\top \V - \I||_F^2$, where $\lambda > 0$ is a chosen parameter, to penalise deviations of $\V$ from being orthonormal. We have found the performance of this method quite insensitive to the setting of $\lambda$, and simply set it equal to the number of data, $n$, since the first term in the objective scales linearly with $n$ as well. Enforcing or encouraging orthogonality of a projection is common in projection pursuit for clustering~\citep{bolton2003,niu2011, hofmeyrSPDC}, and is useful in making the method more robust to poor intialisation. This was not necessary in the classification context since the class labels are known, and hence the estimates of the model parameters are far more stable.

Intuitively this approach can be thought of as asking the question ``Can we find a projection, $\V$, upon which a GMM provides a strong discrimination of the clusters induced by the model?'' Here we mean the clustering induced by the transformed GMM, defined by the parameters $\V^\top \hat\mmu_1, ..., \V^\top \hat\mmu_K$; $\Delta(\V^\top\hat \Sigma_1\V), ..., \Delta(\V^\top\hat \Sigma_K\V)$; and mixing proportions as before. If the parameters of the initial GMM reasonably capture the denser regions of the data, then it is sensible to expect that the additional flexibility in optimising over $\V$ might further enhance this capability. After optimising the projection, as a final step, we re-estimate the parameters of the GMM, but now within the subspace defined by the projection on $\V$.

We experimented with this approach on numerous popular benchmark data sets taken from the UCI machine learning repository~\citep{UCI}. Our main take away from these experiments is that,
especially when the number of dimensions is high relative to the number of observations, frequently the clustering solution changes only slightly from the initial solution. Our understanding of this is that given sufficient degrees of freedom, the application of OPGD serves primarily to further enhance the discrimination of the clusters induced by the initial solution, rather than expose cluster structure in the data which is less identifiable in the full dimensional data set. In a sense OPGD is overfitting to the initial solution.
It is important to note, however, that even when no substantial changes to the initial solution are made, this approach still offers the benefit of a reduced representation of the model, through dimension reduction by projection on $\V$, and also instructive visualisation of the clustering solution which may not be available otherwise. Furthermore, it is encouraging that the majority of the time the changes made to the initial solution, although slight, tend to result in an improved solution.

Table~\ref{tb:cluster} contains a summary of the clustering results from this approach. In each case we used the {\tt R} package {\tt mclust}~\citep{CRANmclust} to obtain the initial GMM solutions, and subsequently applied OPGD to obtain an optimal projection of dimension equal to one fewer than the number of clusters. As in the case of classification, we added very small Gaussian perturbations to the data in some cases in order for a GMM to be fit.
Since the {\tt mclust} implementation uses a random initialisation, we supplied to our method twenty potential initial solutions for each data set. Note that for some of the data sets {\tt mclust}, despite its own random initialisation, obtained the same solution in all twenty replications. In these cases the output of our method was also the same in all twenty. The table reports the performance of the clustering results based on Adjusted Rand Index~\citep{adjustedrand} and Normalised Mutual Information~\citep{nmi}, both multiplied by 100. These are popular evaluation metrics which allow us to numerically assess the similarity between the clustering result and the known class labels. In both cases higher values (close to 100 on our scale) suggest a closer agreement between the clusters and the true classes. For both metrics, and for each data set, we report the average (Avg) performance of the GMM used for initialisation, as well as the average difference induced by the subsequent application of OPGD. In addition we report the single best and worst relative performance of the OPGD output to the initial GMM solution. For example, on the Opt. Digits data set the GMM had an average Adjusted Rand Index score of 49.5 and the average improvement due to OPGD was to increase it to 52.6. The best performance of OPGD was to increase the GMM performance by 10.0, from 54.2 to 64.2, while in the worst instance it decreased the performance very slightly from 42.2 to 41.6. On the other hand, the application of OPGD caused a substantial deterioration in the performance on the Seeds data set, where in every run the Adjusted Rand Index was decreased from 81.2 to 63.8.

\begin{table}[h]
    \centering
    \scalebox{0.8}{
    \begin{tabular}{lccccccc}
          & & \mcl{3}{Adjusted Rand Index} & \mcl{3}{Normalised Mutual Information} \\
         Data Set & $n,d,k$ & Avg & Best & Worst & Avg & Best & Worst\\
         \hline
         Wine & 178,13,3 & 94.9+3.4 & 94.9+3.4 & 94.9+3.4 & 92.8+4.6 & 92.8+4.6 & 92.8+4.6 \\
         Iris & 150,4,3 & 90.4+1.8 & 90.4+1.8 & 90.4+1.8 & 90.0+1.4 & 90.0+1.4 & 90.0+1.4\\
         Vowel & 990,10,11 & 17.5+3.1 & 17.5+3.1 & 17.5+3.1 & 38.6+4.1 & 38.6+4.1 & 38.6+4.1\\
         Satellite & 6435,36,6 & 40.3+2.6 & 40.2+7.2 & 39.6-0.9 & 53.9+3.0 & 59.5+6.5 & 54.0+1.3\\
         Image Seg. & 2310,19,7 & 43.5-0.3 & 43.2+1.4 & 46.0-4.8 & 58.0-0.1 & 57.4+1.3 & 57.4-3.3\\
         Opt. Digits & 5620,64,10 & 49.5+3.1 & 54.2+10.0 & 42.2-0.6 & 65.6+1.8 & 70.0+4.8 & 60.5-0.7\\
         Pen Digits & 10992,16,10 & 57.2+3.8 & 54.2+6.8 & 66.8+1.4 & 75.5+0.5 & 75.2+0.2 & 73.0-1.0\\
         Texture & 5500,40,11 & 88.5-0.1 & 86.6+2.3 & 89.3-1.1 & 93.5-0.3 & 92.4+1.2 & 94.1-1.2\\
         Libras & 360,90,15 & 30.6+1.0 & 32.1+3.4 & 34.0-0.9 & 59.0+1.1 & 60.2+3.3 & 61.3-0.2\\
         Forest & 523,27,4 &  16.8+4.6 & 16.8+4.6 & 16.8+4.6 & 22.8+5.1 & 22.8+5.1 & 22.8+5.1\\
         Yeast & 698,72,5 & 50.4+0.1 & 48.5+2.1 & 55.8-5.1 & 54.5-0.0 & 52.7+1.9 & 56.1-4.5\\
         Glass & 214,9,6 & 10.9+3.6 & -1.8+15.7 & 28.7-9.0 & 28.1+2.7 & 13.2+14.3 & 42.1-5.6\\
         Dermatology & 366,34,6 & 66.5+1.7 & 77.5+6.7 & 45.0-3.5 & 77.2+2.6 & 80.8+7.1 & 72.4-4.0\\
         Seeds & 210,7,3 & 81.2-17.4 & 81.2-17.4 & 81.2-17.4 & 77.1-15.2 & 77.1-15.2 & 77.1-15.2\\
         M.F. Digits & 2000,216,10 & 59.9+3.7 & 70.4+5.4 & 52.0+2.1 & 70.2+3.7 & 71.1+4.7 & 65.2+2.5\\
    \end{tabular}
    }
    \caption{Clustering results from enhancing Gaussian Mixture Models. The values in the table show the performance of the GMMs and the differences in performance resulting from the application of OPGD. The results are based on 20 replications. The ``Avg'' is the average performance of the GMM and the average difference due to applying OPGD, while the ``Best'' and ``Worst'' correspond respectively to the single best and worst performance of the OPGD enhancement relative to the GMM.}
    \label{tb:cluster}
\end{table}

\subsection{Visualisation of Clustering Solutions}

\begin{figure}[h]
    \centering
    \subfigure[Texture: PCA]{\includegraphics[width = .49 \textwidth]{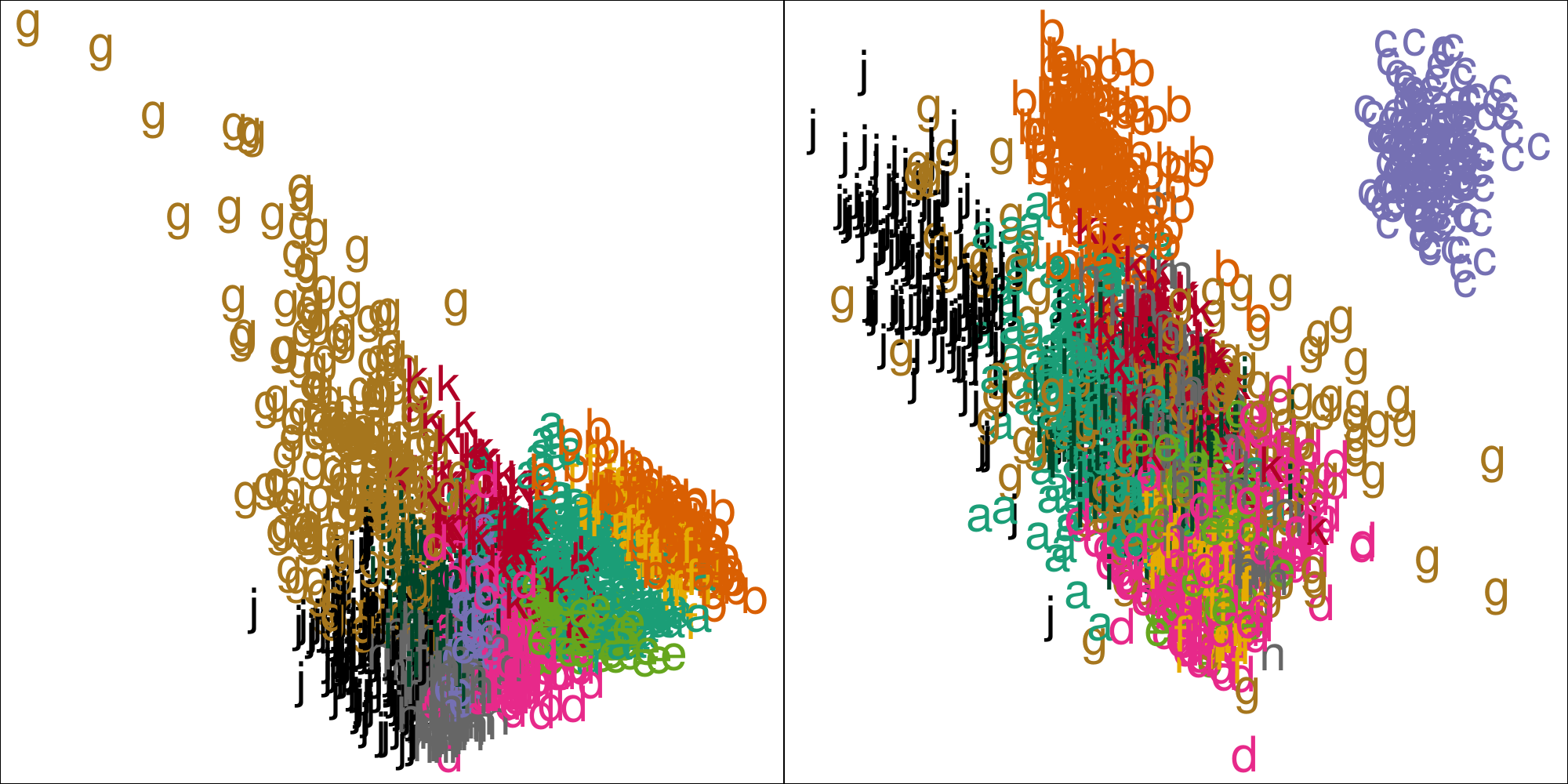}}
    \subfigure[Texture: OPGD]{\includegraphics[width = .49 \textwidth]{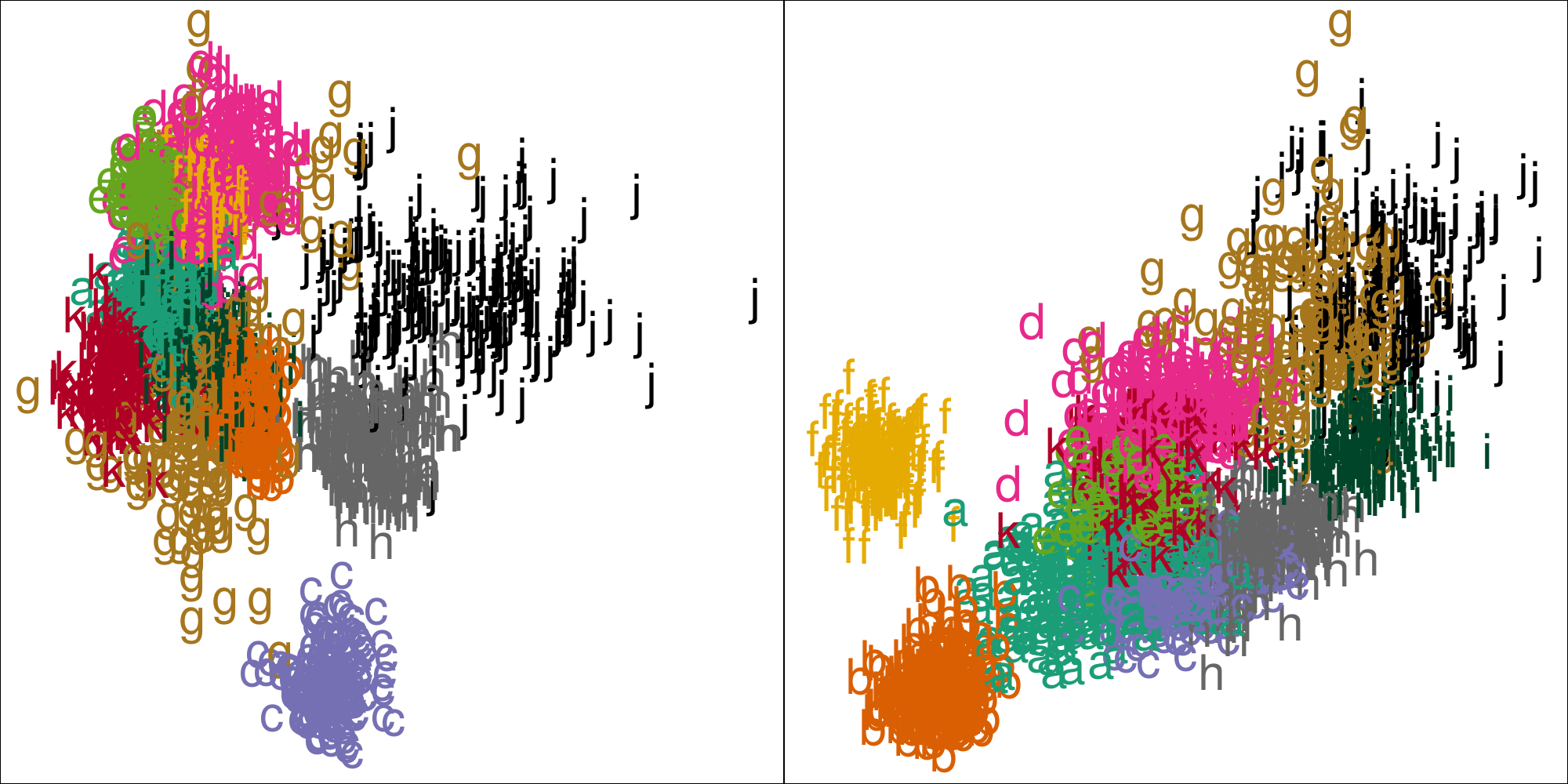}}
    \subfigure[Forest: PCA]{\includegraphics[width = .49 \textwidth]{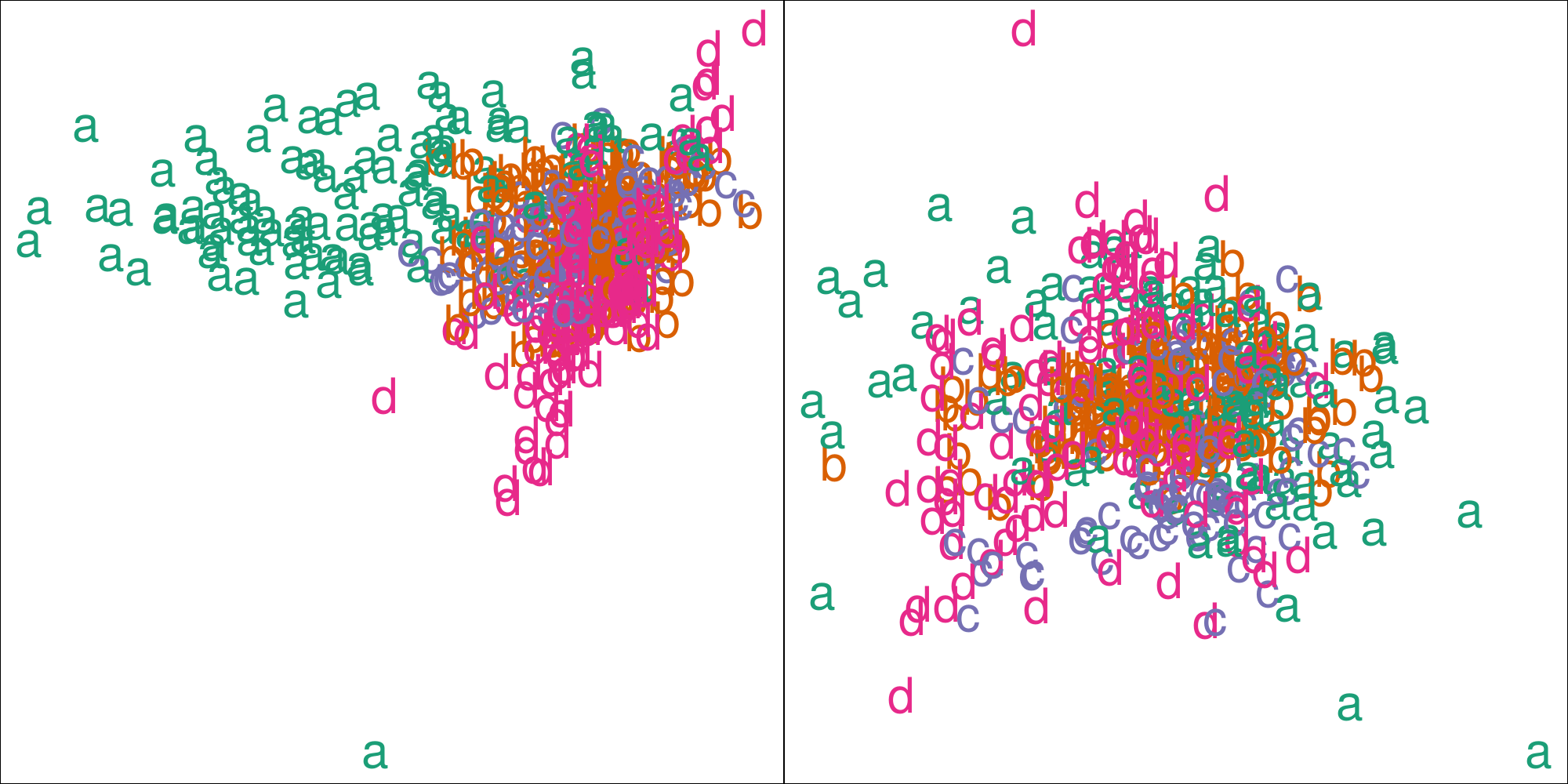}}
    \subfigure[Forest: OPGD]{\includegraphics[width = .49 \textwidth]{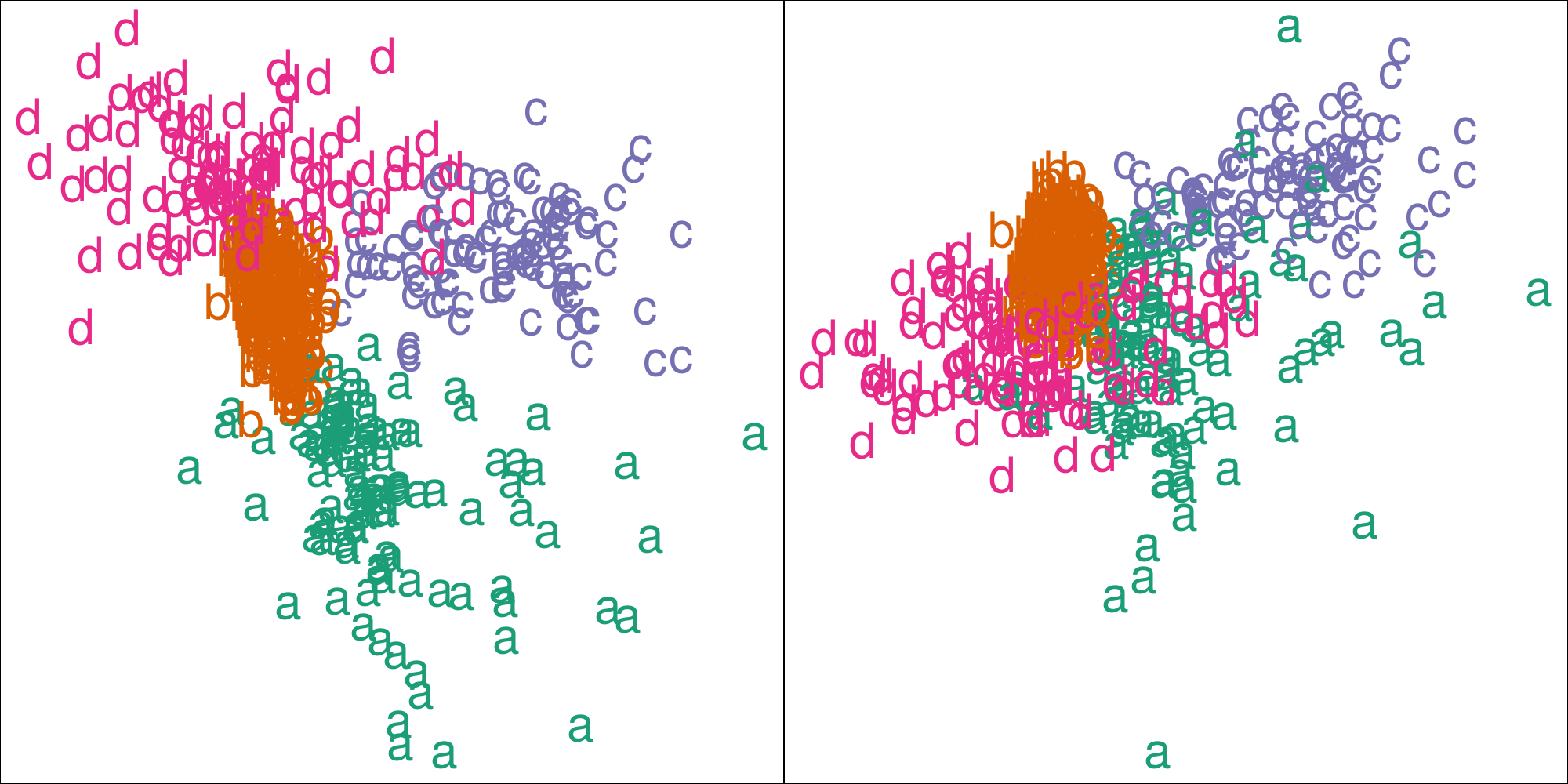}}
    \caption{Visualisation of Texture and Forest data sets through two-dimensional scatter plots. Clusters are far more easily distinguished in the case of the OPGD projections than when using PCA.}
    \label{fig:clust_vis}
\end{figure}

In this section we explore the visualisation of clustering solutions based on the projected data, $\X\V$. As arguably the most popular general purpose method for visualisation, PCA is a natural comparison. Visualisation of clustering solutions is important for, among other things, validation of the solutions obtained. Figure~\ref{fig:clust_vis} shows the examples of the Texture\footnote{For a clearer visualisation we have only included a random subset of 2000 of the total 5500 points in the data set} and Forest data sets. In both cases we provide two 2-dimensional scatter plots for each of PCA and OPGD. In the case of PCA we show the first two and third-and-fourth principal components in the two plots respectively. Here the colours and point characters indicate the clustering based on the initial GMM solution. For OPGD, we do the same for the Texture data set, but since we only seek a 3-dimensional projection for the Forest data set, the two plots in that case are of $\X[\v_1 \ \v_2]$ and $\X[\v_1 \ \v_3]$. In these cases the colours and point characters indicate the final solution. In the PCA plots of the Texture data set one would be hard-pressed to validate more than three of the clusters in the solution. Note that going to lower order principal components does not yield any better visualisation of the different clusters than in the first two pairs depicted in the figure. For the case of OPGD one can easily make out the distinction of at least six different clusters, which agree by eye with how we would intuit clusters in a GMM. Although in the interest of brevity we have not included more than these first two plots, lower order columns in $\V$ can be used to visualise the distinction of the remaining clusters clearly. In the case of the Forest data set, one would not be able to sensibly validate any of the four clusters based on the PCA visualisation, whereas arguably at least three clusters are clearly visible as distinct groups of points in the OPGD projections.

As a final point, in relation to Figure~\ref{fig:clust_vis}, notice that because the objective of OPGD is to maximise discrimination of clusters under a GMM formulation, it tends to lead to solutions, and hence visualisations, in which the clusters appear as roughly having Gaussian, or at least ellipsoidal shapes. Essentially, even when clusters may not be close to Gaussian in the full dimensional space, it is frequently the case that there are projections upon which the clusters are roughly Gaussian. A clear example of such a case is the cluster in the Texture data labelled ``g'' in the PCA plots. These points betray an elongated tail which would not arise if the data came from a GMM with only eleven components. In the case of the OPGD projection, however, all clusters appear as we would expect in a GMM. The distinction of ellipsoidal, or Gaussian clusters is arguably more easily validated by eye than many other non-Gaussian distributions.


\subsection{Underastanding potential failures of OPGD for clustering}

Of the fifteen data sets considered, the performance only deteriorated substantially as a result of the application of OPGD on one; Seeds. However, in a few other cases (e.g., Image Seg. and Texture) very small decreases in average clustering accuracy were also observed.

\begin{figure}
    \centering
    \subfigure[Clustering solution on original data]{\includegraphics[width=5cm]{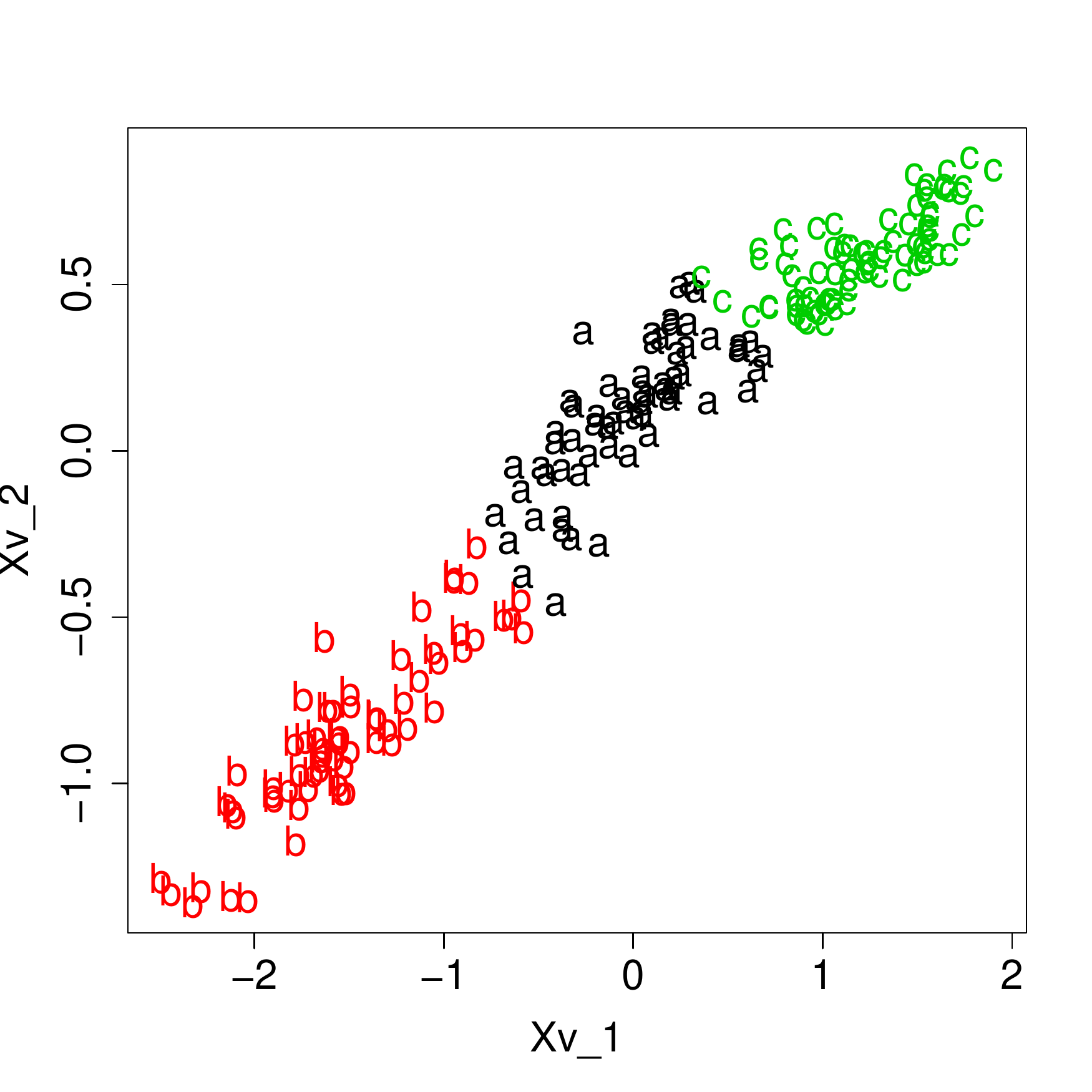}}
    \subfigure[Scree plot of eigenvalues of covariance matrix of data]{\includegraphics[width=5cm]{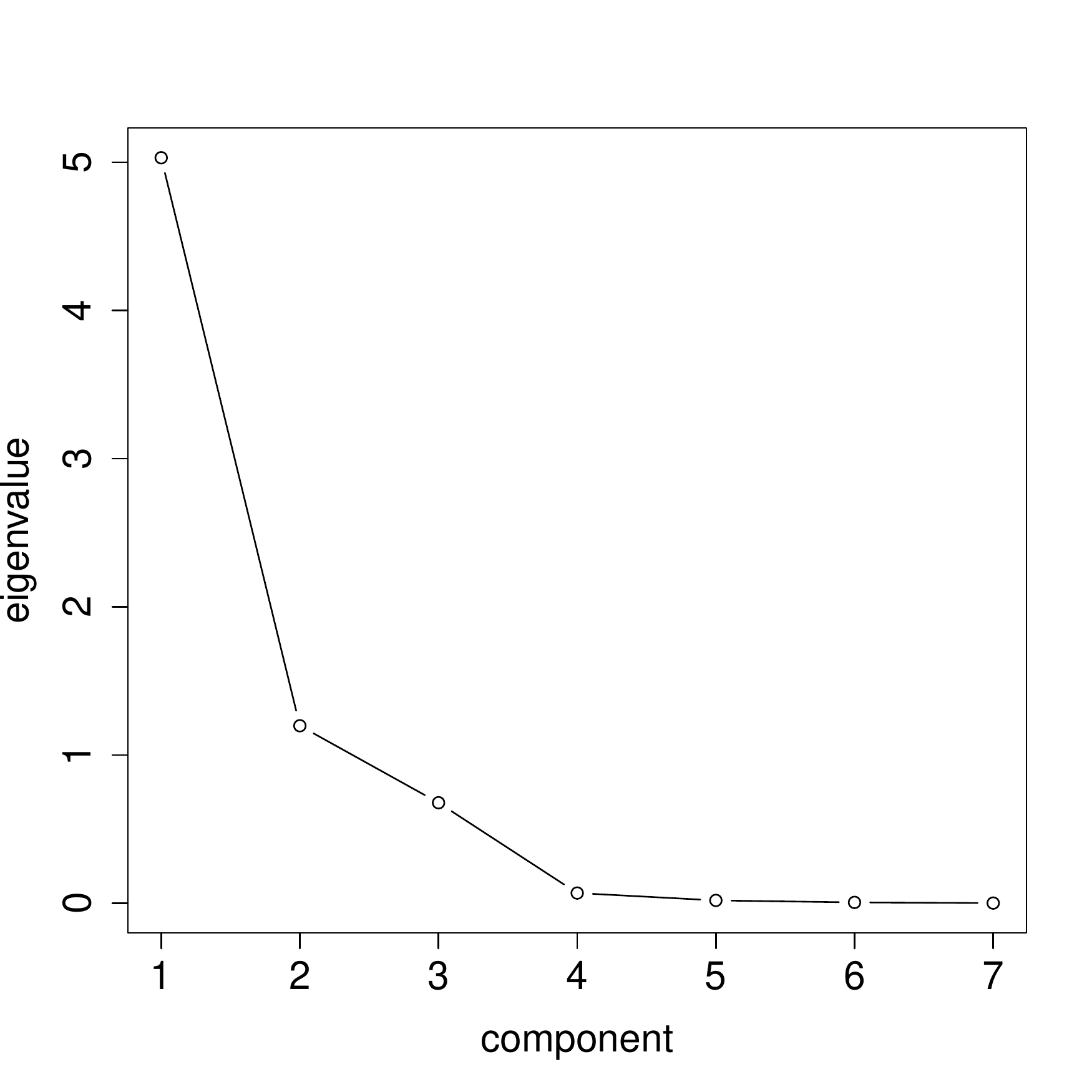}}
    \subfigure[Clustering solution from only first four principal components]{\includegraphics[width=5cm]{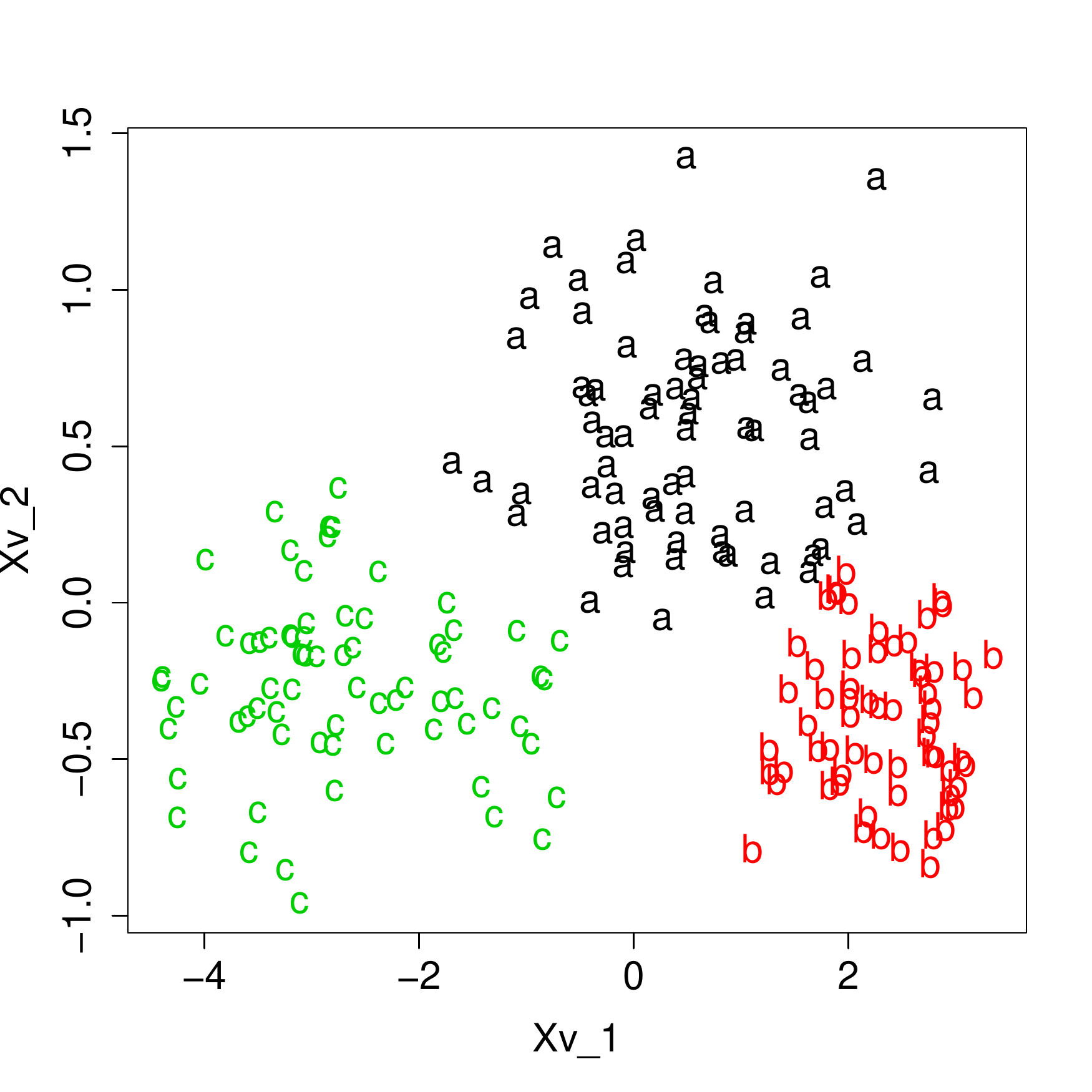}}
    \caption{Seeds data set. (a) shows the OPGD clustering solution of the data projected into the optimal subspace. There is evidence of high collinearity in the data. This evidence of collinearity is corroborated by the very small eigenvalues of the covariance matrix, shown in (b). Removing collinearity by using only the first four principal components of the data leads to a more satisfying solution, shown in (c).}
    \label{fig:seeds}
\end{figure}

It turned out that identifying the potential cause for such a drastic deterioration in clustering performance on the Seeds data set was fairly straightforward. Figure~\ref{fig:seeds} shows plots of (a) the solution obtained by OPGD on the original data; (b) the scree plot of eigenvalues of the covariance matrix; and (c) the solution obtained by OPGD on the reduced data containing only the first 4 principal components. The first plot shows strong collinearity in the data, since despite the fact that the optimal $\V$ is very close to orthonormal we see that the projections $\X\v_1$ and $\X\v_2$ are extremely highly correlated. This collinearity is confirmed by the very small eigenvalues in the covariance matrix of the data. Aside from the standard problems associated with collinearities in clustering,
this example highlights a limitation of our approach in this context. Specifically, this can arise from the combination of a reasonable separation of the means in the initial GMM on a projection, $\v_1$, along which the data are highly correlated with their projection on an orthogonal vector, $\v_2$; and the fact that we fit a diagonal covariance matrix to the clusters within the subspace defined by the projection on $\V = [\v_1 \ \v_2]$. This can be most easily intuited by considering the fact that the optimal projection, $\V$, will be characterised by few points which have relatively high density in more than one of the components of the GMM with parameters $\hat \pi_1, ..., \hat \pi_K$; $\V^\top\hat\mmu_1, ..., \V^\top\hat\mmu_K$; and $\Delta(\V^\top\hat \Sigma_1\V), ..., \Delta(\V^\top\hat \Sigma_K\V)$. In other words, there should be relatively few points on or near the boundaries of the clusters. It should be clear that the proportion of such points is far fewer in Figure~\ref{fig:seeds}~(a) than in Figure~\ref{fig:seeds}~(c). The effect of fitting diagonal covariance matrices in the projected space has the effect of artificially decreasing the density at the boundaries of the clusters in Figure~\ref{fig:seeds}~(a) since the density is spread almost equally in the direction along which the data lie and the orthogonal direction. By removing the collinearity in the data the solution is vastly improved, Figure~\ref{fig:seeds}~(c). In this case the performance is similar for the initial solution and the adjusted solution after applying OPGD, with both metrics increasing slightly from the initial values.

We note that it is likely not necessary to always check for such collinearity prior to applying OPGD, as the scenario which led to the poor performance on the Seeds data sets was very specific. However, for completeness, we explored the alternative of applying {\tt mclust} and then OPGD again to all the data sets considered before, except now only the leading principal components which constituted at most 99.9\% of the total data variation were included. The threshold of 99.9\% is arbitrary, and was chosen only for a preliminary investigation into the effect of removing only very small variance components. In none of the other cases did this substantially affect the relative performance of the OPGD enhancement when compared to the GMM initialisation.

\section{Discussion} \label{sec:conclusions}

In this paper we explored the problem of finding optimal projections for discriminant analysis in which each class is endowed with a multivariate Gaussian density. By carefully addressing the differentiation of the objective function we were able to optimise the classification likelihood directly with gradient based optimisation techniques. We found that this approach is very successful in obtaining accurate classification models when compared with other popular Gaussian discriminant models. A simple modification of this objective allowed us to also address the problem of enhancing Gaussian Mixture Models for clustering. We found this approach to consistently offer modest improvements in clustering quality, while simultaneously providing a reduced model formulation through dimension reduction, as well as instructive visualisations for cluster validation and knowledge discovery.

\bibliographystyle{plainnat}
\bibliography{OPGD}


\end{document}